\documentclass[12pt]{amsart}%
\usepackage{graphicx}
\usepackage{amsmath}
\usepackage{amsfonts}
\usepackage{amssymb}
\usepackage{fancyvrb}
\usepackage{longtable}
\usepackage{setspace}
\usepackage{etoolbox}
\AtBeginEnvironment{footnote}{\singlespacing}
\AtBeginEnvironment{abstract}{\singlespacing}

\providecommand{\U}[1]{\protect\rule{.1in}{.1in}}
\usepackage{color}

\title{Equitable Longevity Risk Sharing \\ or, the {\em raison d'etre} for a \\ First Nations Pension Plan}

\author{Moshe A. Milevsky, Thomas Salisbury \& Robyn Allen}
\thanks{Milevsky is a Professor of Finance and CIT Chair in Financial Services at the Schulich School of Business, as well as the Executive Director of the IFID Centre. Salisbury is Emeritus Professor of Mathematics \& Statistics. Allen is with Osgoode Hall Law School, all three are at York University in Toronto, Canada. The authors acknowledge funding from the CIT Chair  (Milevsky), The IFID Centre (Milevsky), SSHRC (Milevsky, Allen) \& NSERC (Salisbury). The contact author can be reached at milevsky@yorku.ca. The authors would like to acknowledge helpful comments and suggestions on an earlier draft, from An Chen, Edward Doolittle, Irene Henriques, Madeleine Salvadori and Colin Scott.}

\date{Version: 27 November 2025 (submitted).}

\begin{document}

\begin{abstract}
We investigate the extent to which groups with elevated mortality rates {\em ex ante} might opt out of guaranteed national pensions in favour of demographically aligned plans, which we label {\em equitable longevity risk sharing} (ELRiS) pools, even if this involves accepting some idiosyncratic risk. Technically, this paper develops a stochastic model of retirement income within an ELRiS structure that is calibrated to equate the discounted expected utility of a guaranteed national pension. The practical motivation for developing this alternative is that working members of First Nations peoples of Canada: (1) experience much higher mortality rates than average over their entire life cycle, and (2) some are actually allowed by current legislation to opt out of the Canada Pension Plan (CPP).  We then demonstrate that under reasonable economic preferences and parameters, a sub-group with a 10-year life expectancy gap relative to the population could attain equivalent lifetime utility by contributing a mere {\em two-thirds} to the plan, even if they were pooled with only 30 members. For a longevity gap of 20 years, such as between an Indigenous male versus a non-Indigenous female, the contribution rate falls to less than {\em a third.} The difference between the statutory and mandatory contribution rates to a guaranteed national pension and those needed within these self-sustaining pools is an implicit subsidy from Indigenous to non-Indigenous.  From a policy perspective, this paper aspires to jump-start a conversation that sparks a change in a {\em status quo}, which is obviously unfair and inequitable.
\end{abstract}

\maketitle

\clearpage

\section{Motivation \& Background}
\label{sec:motivn}

One of the foundational axioms within the field of pension economics \& finance is that (risk-averse, rational, utility-maximizing, consumption-smoothing, etc.) retirees are better off sharing and pooling their financial resources while decumulating. Most economists encourage the young to join (large) groups of eventually-to-be elderly workers in which {\em longevity risk} can be diversified away, others' mortality credits can be harvested, and life annuities are assured for all.  In fact, it is the mathematical {\em Law of Large Numbers} -- rather than the economics of professional money management -- that remain the {\em raison d'\^etre} for large guaranteed Defined Benefit (DB) pension arrangements, versus individually managed Defined Contribution (DC) accounts.\footnote{There are of course many behavioral and financial literacy arguments (and even more papers) in favour of institutional DB over individual account DC. Still, the main economic difference between the two extreme arrangements is the \underline{cheaper} guaranteed income that stems from the longevity risk sharing.} The risk sharing argument is especially important to emphasize in a world where investment management fees on individual accounts, funds and products have plummeted, thus killing the {\em scale economy} arguments behind larger (and larger) pension plans.  Actuaries will remind policymakers that individuals who decumulate alone lose the actuarial mortality credits compared to those who decumulate in a group. So argue the pension economists as well, such as the textbook arguments made by Bodie (1990), Blake (2006), or Barr \& Diamond (2009), to name just a few of the well-known sources. 

\vspace{0.1in}

The vast literature on this topic also falls under the umbrella of the {\em annuity puzzle}, predicated on the same arguments.\footnote{For recent work, see: Arandjelovi\'c, A., \'Ciri\'c, M., \& Nedeljkovi\'c, M. (2023), Bello, P., Brugiavini, A., \& Galasso, V. (2024), B\"utler, M., \& Ramsden, A. (2024), Charupat, N., \& Kamstra, M. J. (2024), Clark, R. L., \& Mitchell, O. S. (2024), Fong, J. H., \& Li, J. (2022), Hagen, J. (2022), Jang, C., Clare, A., \& Owadally, I. (2022), Lau, S.-H. P., \& Zhang, Q. (2023), Pashchenko, S., \& Porapakkarm, P. (2022), Searle, P., Ayton, P., \& Clacher, I. (2024), Unger, F., Steul-Fischer, M., \& Gatzert, N. (2024). All papers and authors in this exhaustive list would agree that some form of annuitization is socially desirable and that DB pensions offer this benefit. Maurer, Mitchell, Rogalla, and Kartashov (2013) is an early article that notes the benefits of introducing these arrangements into institutional plans.} Namely, it's better to annuitize {\em ex ante}, even if {\em ex post} there will be winners who live to consume for a long life at the expense of longevity losers who don't.  Now, admittedly, there is an awareness that a certain amount of heterogeneity is degrading the benefits, but these cross-subsidies -- i.e. money from the dead going to support the living -- are accepted as the cost of maximizing social welfare. The pivotal pension idea imported from insurance economics is that those with a lower life expectancy can still benefit from the arrangement, even if they are forced to pool resources with those who have better prospects. This is no different than home, car, health or even pet insurance. See, for example, Varian (1992), or Eeckhoudt, Gollier \& Schlesinger (2011), as long as the `loading' isn't too high, and for this we point to the early work by Mitchell, Poterba, Warshawsky, and Brown (1999), who make this exact case. Annuity loadings aren't too high. 

\vspace{0.1in}

We don't mean to imply these transfers are inefficient or unfair, and may be unintended though disproportionate consequences of large-scale societal pooling and risk sharing. As just one example of the awareness of the cross-subsidies when it comes to pensions and annuities, Milevsky (2020) quotes the early work by Brown (2003), who writes: ``The insurance value of annuitization is sufficiently large that, relative to a world with no annuities, all groups can be made better off through mandatory annuitization.'' An article by Arapakis, Wettstein, and Yin (2023), perhaps motivated by concerns of distorting cross-subsidies identified by Brown (2003), argues in a similar vein. They examine the insurance value of (US) social security by race and socio-economic status, since it is well known that mortality rates for Black households (in the US) are higher than the mortality rates for White households. {\em Prima facie}, the former should lose at the expense of the latter. Yet, Arapakis, Wettstein, and Yin (2023) conclude ``Black households derive more longevity insurance value'' from this program, which pools all heterogeneous Americans together in one large pool, compared to their ``White counterparts.'' Brown (2003) also wrote about U.S. social security and annuitization more specifically, that ``When measured on a financial basis, these transfers can be quite large, and often away from economically disadvantaged groups and towards groups that are better off financially.'' The awareness is clearly there. More recently, Thalagoda et al (2025) examine pension and annuity transfers via a Gini-based assessment framework and state up-front that: ``Pooling retirees with heterogeneous longevity and wealth profiles can lead to unintended wealth transfers that benefit some groups at the expense of others.'' Reviewing the pension and annuity literature one finds that these discussions are often accompanied by reassuring statements that a progressive income tax system ensures that any pre-tax transfers from the dis-fortunate to fortunate (bad), are reversed on an after-tax basis (good.)\footnote{This is a rather contentious matter, and we simply reference the literature from Coronado et al (2011), to Groneck and Wallenius (2021) for an overview.}

\vspace{0.1in}

The challenge with this mindset — or perhaps better phrased as what is absent from the literature — is that it does not offer guidance on the precise point at which a very large longevity risk-sharing pool no longer makes equitable sense and should be dismantled. More specifically, the literature has not thoroughly investigated the longevity thresholds at which groups with higher mortality rates should secede and create their own 'sharing republics.' Why would they bother changing the {\em status quo?} Answer: This is not about economic {\em segregation} but rather financial {\em savings.} The claim is these subgroups could contribute {\em less} to their pensions, and obtain the same level of (utility of) retirement income. Again, our aim is not to suggest that national pension systems be divided along politically fraught ethnic or cultural lines. Rather, we aim to explore whether risk-sharing structures could be designed for smaller groups with demonstrably different longevity profiles, in a more equitable manner. This aligns with a sub-stream in the actuarial literature that is investigating similar questions.

\vspace{0.1in}

The motivation for asking and investigating these sorts of research questions {\em now}, and especially in 2025, is as follows. One of the most troubling statistics in demography and actuarial science is the significantly higher mortality rate among Indigenous populations around the world. The Indigenous peoples of Canada, for example, face a life expectancy gap of 10 to 15 years below the national average, depending on where they live, and this gap persists across time and geography. It is even worse for those who live on traditional reserves. Two scholars who have documented the exact numbers are Donna Feir and Randall Akee. In a (2019) paper, they write: ``Overall, the mortality rates for the on-reserve population tend to be almost always twice that or more of the Canadian averages for males across most of the five-year age groups.'' Further, ``The rate for females on reserves is often triple to quadruple the Canadian average at many ages.'' This isn't merely abnormally high teenage or infant mortality rates. Rather, it is a gap across the entire mortality curve. The historical background -- and some reasons for this -- will be discussed in the next section. Here, we note that Cooke and Shields (2024) offer a systematic review of the literature on health and mortality outcomes, and argue that anti-Indigenous racism in Canada reduces access to healthcare and is thus one of the transmission channels for the higher mortality. In yet another identified factor, Shapiro et al (2021) argue that substandard housing leads to the spread of infectious diseases among the Indigenous peoples of Canada, which increases mortality across the entire age curve. 

\vspace{0.1in}

Doubling or tripling a mortality hazard rate might sound rather abstract and not very meaningful in the context of pensions and retirement. So, to put these numbers in perspective consider the following. Two distinct demographic groups save the same dollar amount each year in one large joint pool. Though both face a risk of early death and not retiring, the second group is assumed to have double the mortality rate of the first. Now, doubling a (yes) constant mortality rate from say $0.01$ to $0.02$ per year, over $30$ working years, the survival probability drops from $e^{-(0.01)30}$ to  $e^{-(0.02)30}$. That is from 74\% for the first group to $54.9\%$ for those with double the mortality rate in the second group. It collapses to $40.7\%$ for those with triple the mortality rate. That is the order of magnitude. 

\vspace{0.1in}

To be very clear, while this health and longevity disparity that we are shining a strong light on raises significant and pressing concerns for public policy and social justice -- and has been lamented, documented, and explored in papers such as the above-noted work by Feir and Akee (2019), Cooke and Shields (2024) or Shapiro et al (2021), its implications for retirement provision and pension design remain under-explored. In fact, the entire topic of personal wealth management and finance for the Indigenous tends to get little attention by the standard academic literature. It is particularly salient and relevant in Canada, where, though not well known or advertised, {\em some} working members of First Nations can actually opt out altogether from the Canada Pension Plan (CPP), a mandatory second-pillar pension for all Canadians outside Quebec. This feature is unique to Canada. No other country allows its Indigenous population to opt out of the national pension. Given the little attention that has been paid to this matter, we will carefully explain the source of this `option' to opt out and some of the data available on those who don't. In fact, to preview some of our efforts, the authors had to file a formal freedom of information request to obtain data on the number of First Nation members who participate in CPP -- but may not have to. 

\vspace{0.1in}

This paper and the underlying research agenda provide some context and technical insight to the question: {\em At what point could there be a separate First Nations Pension Plan and what should it look like?} More broadly, we examine whether those with elevated mortality might rationally shun large guaranteed national pensions in favour of smaller, demographically aligned pools -- even if this involves accepting some idiosyncratic risk. In other words, this paper also (1.) offers an alternative to national pensions for individuals with noticeably higher mortality rates; (2.) explains precisely how {\em equitable longevity risk sharing} (ELRiS, rhymes with Elvis) pools might work; and (3.) using some reasonable parameters for the contribution and replacement rates of the CPP, illustrates how much participants could save by seceding from the national plan.

\subsection{Outline of the Paper}
\label{subsec:outline}

The remainder of this paper is organized as follows. The next section (\#\ref{sec:CPP}) offers a brief overview of the Canada Pension Plan (CPP). The subsequent section (\#\ref{sec:FN}) provides some context, background, history and the regulations pertaining to First Nations and the CPP. Section \#\ref{sec:model} develops a normative model in which we compute how much First Nations members -- or any group with impaired mortality -- would gain from not joining the national pension plan, and instead share longevity risks among themselves. Then, in section \#\ref{sec:results} we provide and interpret the numerical results. The paper concludes in section \#\ref{sec:conclusion} with a high-level summary and some policy implications. All non-essential derivations are placed in a technical appendix. Although there are a few subtle mathematical issues that push the envelope of existing theory, we don't want to distract the primary audience for this paper, pension policymakers and financial economists, from the main thrust of our argument.

\section{The Canadian Pension Plan (CPP)}
\label{sec:CPP}

The most recent scholarly article on the structure of both Federal and Provincial income support programs in Canada is by Tammy Schirle and is cited as Schirle (2024) in the bibliography. Moreover, we rely on some (though not all) of her parameters for the calibration of our model in section \#\ref{sec:results}, and here we offer a very high-level summary of (only) the Canada Pension Plan (CPP), not Old Age Security (OAS), the Guaranteed Income Supplement (GIS), or any provincial programs. The CPP is a mandatory public pension system that provides retirement, disability, and survivor benefits to workers and their families across most of Canada. The province of Quebec has its own parallel version, the Quebec Pension Plan (QPP), with similar rules -- although for our purposes, we will point out that they use Quebec mortality experience to determine benefits. That {\em is} an example of a smaller pool. Likewise, and we tread carefully here, the Canadian Province of Alberta has recently threatened to do something similar to Quebec. In fact, provinces can legally secede from the CPP, with the only question being what fraction of the assets they can reasonably demand.\footnote{We reference the article by Bob Baldwin (2024), who concludes: ``The creation of an APP will involve taking on a good deal of risk for relatively small gains that are uncertain over the long term.'' In contrast, Alberta Premier Danielle Smith is in favour of removing Alberta from the Canada Pension Plan and starting its own Alberta Pension Plan. The current authors have nothing to contribute to that debate. This paper is focused on First Nations as a group, regardless of province. We should note that Nunavut is the jurisdiction with the highest percentage Indigenous population, and while the Indian Act does not exempt the Inuit from taxes, the arguments we make in this paper suggest that an Alberta-style push for secession from CPP could actually make more sense there. Likewise, The Northwest Territories (NWT) has the highest percentage of First Nations residents, so if a territorial government was ever going to advocate for a standalone scheme, it might be there.}

\vspace{0.1in}

Be it CPP or QPP, these large, diversified pension plans are funded through contributions made by both employees and employers, as well as by self-employed workers, and are designed to replace part of a worker's income in retirement or in the event of disability or death. CPP/QPP is funded on a pay-as-you-go basis with significant but partial funding. Employees and employers each contribute a set percentage of the worker's pensionable earnings up to a yearly maximum, while self-employed individuals pay both shares. So an employee might be required to contribute 6\% of their wage (for example), but someone who is self-employed and pays themselves via a corporation would have to contribute 12\%. We will return to this doubling later. Either way, these contributions are collected via the Canada Revenue Agency (CRA) infrastructure, as part of the tax system, and then invested by the Canada Pension Plan Investment Board (CPPIB). This professional body manages the fund to ensure long-term sustainability by targeting a real rate of return. To calibrate one's intuition (only), consider the employee who contributes 6\% of their wages to a stylized CPP, with the employer matching this with an additional 6\%, resulting in a total contribution of 12\% to the plan over a working life of, say, 40 years. At retirement, the plan generates an annual income of approximately one-third of those wages, for life. 

\vspace{0.1in}

Workers can begin collecting CPP retirement benefits as early as age 60 or as late as age 70. The typical or common age is 65, but starting earlier results in a permanent reduction, while delaying increases the monthly payment. Generally speaking, the plan is actuarially biased, particularly benefiting those with higher life expectancy and no liquidity constraints, by encouraging later rather than earlier claims. This design flaw is (also) not something we address in this paper; instead, we refer to Charupat and Parlar (2017) for an analysis of that financial tradeoff, or the more recent work by Macdonald (2024).  Either way, early or late, the pension benefit amount depends on an individual's contribution history, average pensionable earnings, and the age at which benefits are started, which is, again, under the retiree's choice. At maximum, CPP/QPP currently -- that is, in 2025 -- replaces about one-quarter of average earnings, though most people receive less than the maximum due to contribution gaps or lower earnings. However, and this is very important for our simplified modelling, this ratio (25\%) is engineered to increase over time to approximately one-third (33.33\%). In other words, the plan aims to replace a third of Canadian's workers' income (subject to caps, up to limits.) This is known as the CPP enhancement, and here we refer to  Macdonald (2019) for more on that aspect of the CPP. Beyond retirement income, true (versus our stylized) CPP also provides: disability benefits for contributors who are unable to work due to severe and prolonged disability; survivor benefits for spouses/common-law partners and dependent children of deceased contributors; and death benefits, which are one-time payments to a deceased contributor's estate.

\vspace{0.1in}

Finally, as far as the parameters are concerned -- and once again we refer to Schirle (2024) for some of the nuances -- in 2025, the CPP contribution rate for employees and employers is 5.95\%, up from 5.25\% in 2020, as part of the above-noted multi-year CPP enhancement that began in 2019. Back in 2020, the Year's Maximum Pensionable Earnings (YMPE) was \$58,700, compared to \$68,500 in 2025, meaning the maximum annual contribution has grown from about \$2,898 per person in 2020 to roughly \$3,867 in 2025. Once again, employers match, and the self-employed pay both shares. These staged increases in contribution rates and earnings ceilings were designed to gradually strengthen the plan, raising CPP's income replacement rate from (the above-mentioned) 25\% to about 33\% and ensuring future retirees receive higher, more sustainable benefits. The true CPP is converging towards the stylized CPP we harness in this paper.

\vspace{0.1in}

Note also that the CPP payments are adjusted annually for inflation. So, for example, in 2024, all payments were increased by 4.4\% nominally. There is also a mechanism for, in effect, adjusting past pensionable earnings (which determine the target the income replacement percentage will be applied to) for inflation. This is why our model (in section \ref{sec:model}) and parameters selected (in section \ref{sec:results}) will be in real inflation-adjusted terms.

\section{First Nations, Pensions \& Longevity Risk}
\label{sec:FN}

In this section, we explain why a small group of Canadians have been granted the option to opt out of the CPP, when all other Canadians are obligated to contribute to this plan or fund if they are in the labour force. A brief examination of Canadian history as well as the \textit{Indian Act} will help (1.) appreciate the reasons for this exemption, and (2.) understand the root causes of the elevated mortality rates, which are the driving motivation for this paper. For context, we begin with a brief overview of population statistics.

\vspace{0.1in}

According to the (most recent) 2021 census, Canada's population consists of approximately 36,991,981 people. Of those, a mere 5\% identify as Indigenous. At 1,807,250, the Indigenous population of Canada is small relative to the non-Indigenous population. However, the same census data indicates that the Indigenous peoples of Canada continue to grow steadily. Indigenous people(s) have played a foundational, albeit complicated, role in the history and creation of Canada. The Indigenous population of Canada comprises of three distinct and culturally diverse groups of people. They are: First Nations, M\'etis, and Inuit. Oddly enough, though they are all Indigenous to Canada, the rules that apply to each group pertaining to pensions and retirement income are different. At 1,048,400, First Nations represent more than half of the Canadian Indigenous population. This is followed by the M\'etis and then Inuit at 624,220 and 70,540, respectively. 

\vspace{0.1in}

First Nations are unique in that they are the only Indigenous group in Canada to be governed by the \textit{Indian Act}. Drilling down to a deeper level, more than half of First Nations members are also registered under the \textit{Indian Act} and of those who are registered, approximately 311,895 live on-reserve. Now to the key fact here. There are certain exceptions under the \textit{Indian Act} that relieve certain First Nations individuals of the obligation to pay Canadian income tax, which is linked to the exemption for some from contributing to the CPP, but they can contribute if they want, subject to certain conditions. More clarity in a moment. But first, here is a bit more of the background that might help explain the context within which this paper resides.

\vspace{0.1in}

First Nations,  historically referred to as ``Indians'' have inhabited North America since time immemorial and have existed long before Europeans settled the land.  More broadly, today, there are hundreds of First Nations communities across the country, each with their own unique stories, histories, and traditions. First Nations communities were self-governing; they made use of the land, which in turn provided all the necessities of life. But with the arrival of European settlers, they experienced a dramatic decline in sovereignty, which had obvious implications for health and welfare. The land inhabited by First Nations quickly became a landscape for war between colonial powers, namely the British and French. The objectives were wealth, territory, and colonial supremacy. Despite serving as critical allies and negotiating multiple peace and treaty agreements with settlers, First Nations people did not regain the level of autonomy they had pre-contact. With the cessation of colonial wars, the British and eventually the Canadians turned their attention to what at the time was labeled, the ``Indian Problem''. In 1876, the government of Canada created the \textit{Indian Act}, which granted authority over ``Indians'' to the government.\footnote{We do not even pretend to offer a comprehensive review of the Indian Act, and the need to change or perhaps eliminate the {\em Act}, but refer royal commission report by Giokas (1995), and to Richmond \& Cook (2016), as well as McLean, et al (2002) for health matters. See Colbourne et al (2024), for a recent survey of the legal, political, and socio-economic forces that undermine Indigenous entrepreneurial activity and well-being.}

\vspace{0.1in}

The purpose of the \textit{Indian Act} was two-fold: to exert control over First Nations and to assimilate them into Canadian society. First Nations registered under the \textit{Indian Act} acquired ``Indian Status''. However, the same act also stipulated that ``Indians'' were effectively ``non-people''. Until 1951. A ``person'' was defined in the act as an individual other than an ``Indian''. The act stipulated that ``Indians'' could become ``people'' by means of voluntary or involuntary enfranchisement. In other words, voluntarily or involuntarily relinquishing ``Indian Status''. For example, an ``Indian'' could be enfranchised by serving in the Canadian Armed Forces, obtaining post-secondary education, or marrying the wrong person. Indeed, many First Nations women were stripped of their ``Indian Status'' because the act denied women and their children their status if they married a non-``Indian'' or non-``Treaty Indian''. As summarized by the Native Women's Association of Canada, ``enfranchisement created an either-or scenario for Indigenous people: you could be `Indian' or you could be `Canadian'.'' The \textit{Indian Act} is also responsible for creating the reserve system and residential schools.\footnote{The residential school system separated thousands of First Nations, M\'etis, and Inuit children from their families. Children were sent to residential schools where they were forbidden from practicing their cultures or speaking their native languages. Students were routinely subjected to horrific physical, emotional, and sexual abuse, often at the hands of religious authorities. Shockingly, the last residential school in Canada did not close until the late 1990's. the \textit{Indian Act} has since been amended but is still in force today.} 

\vspace{0.1in}

With this context in place, and specifically for pension researchers, the \textit{Indian Act} is particularly notable because of the embedded statute related to income tax. The tax exemption in Section 87(1) of the \textit{Indian Act} generates the optionality to participate in the CPP.  Accordingly, members of First Nations who are registered under the \textit{Indian Act} and who have ``Indian Status'' can avail themselves of the tax exemption on property and income if they can prove that it is sourced to reserve lands. \textit{Moreover, the act creating the CPP absolves First Nations individuals who meet the tax-exemption criteria from having to make contributions to the CPP.} Instead, they could in principle save those contributions (e.g. 6\%) and invest alone or with others in a pooled fund of their choice. To be clear, though, members of First Nations working off-reserve must contribute to the CPP. So, from a practical perspective, our audience for this paper are those working on reserve, though we should note that the surface area of all reserve land in Canada is approximately 8 million acres. 

\vspace{0.1in}

This, again, is critical because the life expectancy of First Nations is approximately 10 years less than that of non-Indigenous Canadians. Beyond the articles cited earlier, a study conducted by Statistics Canada analyzing mortality rates and causes of death between 2006 and 2016 revealed that First Nations living on and off-reserve succumb to the following causes of death at higher rates than non-Indigenous Canadians: homicide, suicide, accidents, nephritis, nephrotic syndrome and nephrosis, chronic liver disease and cirrhosis, chronic lower respiratory diseases, influenza and pneumonia, and diabetes. In particular, causes of death related to diseases of the heart, cerebrovascular diseases, and cancer are higher for First Nations living on-reserve than for non-Indigenous Canadians. 

\vspace{0.1in}

\begin{figure}
\vspace{-0.1in}
\caption{Mortality of First Nations people of Canada relative to non-Indigenous}
\label{fig1}
\vspace{-0.1in}
\begin{center}
\includegraphics[width=1.0\textwidth]{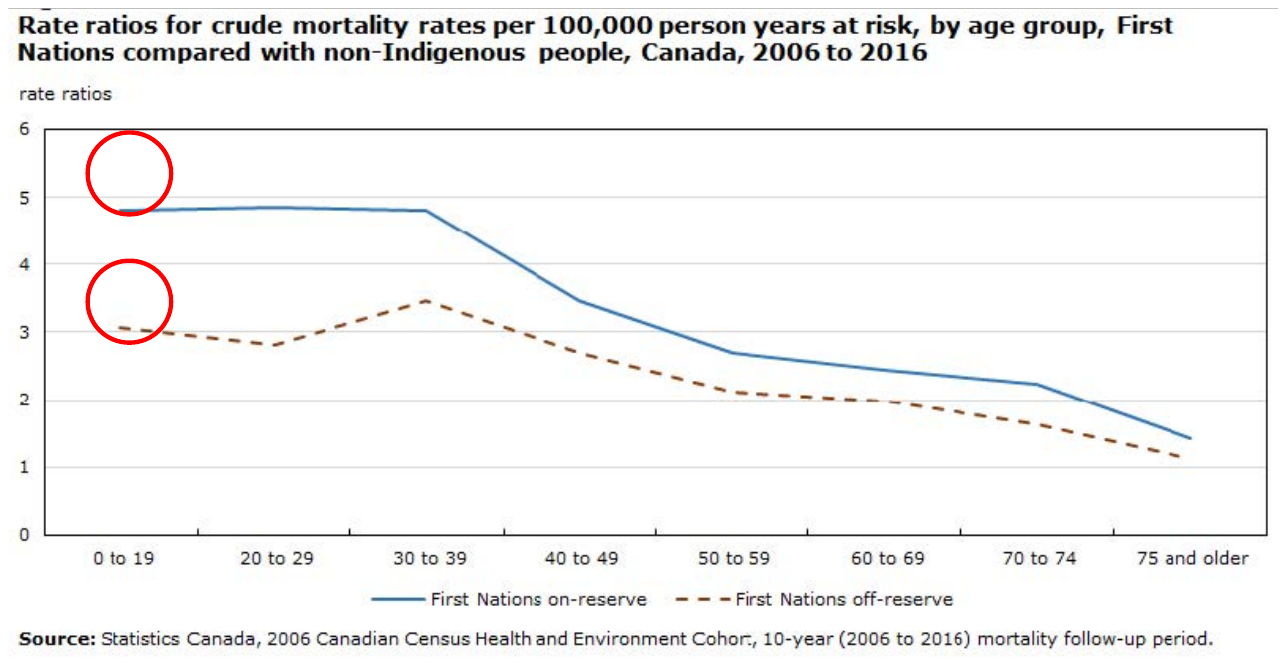} 
\end{center}
\end{figure}

\begin{center}
{\bf [Figure \#\ref{fig1} goes here]}
\end{center}

\vspace{0.1in}

Beyond the Gompertzian or aging-related aspects of mortality, suicide rates among First Nations youth are 5-6 times higher than non-Indigenous youth. Similarly, the rate of infant death among First Nations is more than twice as high as compared to non-Indigenous Canadians. Once again, First Nation mortality rates are agonizingly high across all age groups, including those impacted by retirement planning. Figure \#\ref{fig1} is a visual indication -- using official statistics compiled and displayed by the Government of Canada -- of how mortality rates are higher across the entire mortality curve, although they do converge at higher ages as one might expect from the compensation law of mortality. See the most recent work by Gavrilov and Gavrilova (2024) -- though they have been at this for many decades -- for more on this `law' and why, at very advanced ages, all mortality rates (may) converge. See, also Milligan and Schirle (2021) for more on Canadian mortality trends as they relate to pensions. 

\vspace{0.1in}

If, on average, adult First Nations are dying approximately 10 years before non-Indigenous Canadians, this begs the following question: Why should First Nations members who are exempt from making CPP contributions do so, if they are effectively subsidizing the retirement income of others who have better longevity prospects? Stated differently, First Nations members are not in a position to realize the same internal rate of return (IRR) on their CPP investments, as non-Indigenous Canadians.  Their IRR's will be worse. 

\subsection{But are any of them actually contributing?} 

This information is not (easily or) publicly available on the various agency websites. To answer this particular question, the authors made a formal request under the Canadian {\em Access to Information Act}. This query -- formally an ATIP request --- allows any and all Canadians to access information from federal government agencies, that is not otherwise available to the public\footnote{The original ATIP request was filed on the 7th of May, 2025, and after much back and forth, we received what we believe is the correct disclosure or information on the 21st of July 2025. Note that the data we were initially provided under the ATIP request was calculated by CRA using CPP credits versus deductions. Moreover, on this point, there is a dearth of information about CPP planning made available to the indigenous. Casual empiricism suggests few members of First Nations are aware of this optionality}. We obtained the following reply. According to the Canada Revenue Agency (CRA), a total of 90,566 members of First Nations with net tax-exempt income, made contributions to the CPP or QPP in 2023. In the same year, a total of 3,068 First Nations with net tax-exempt and self-employed income also made contributions, which means they had to double the contribution rate (employer and employee). Table \#\ref{tab:mydata} displays the actual response which is reproduced verbatim, to the (ATIP) request, including the precise format in which the data was transmitted. Clearly, a substantial number of First Nations members -- who do not have to contribute to the CPP by law -- do so in practice. The questions of {\em why} is thus even stronger.

\vspace{0.1in}
\begin{center}
{\bf [Table \#\ref{tab:mydata} goes here]}
\end{center}
\vspace{0.1in}

In practice, if an on-reserve employer chooses to participate in the CPP, employees are also obligated to participate and make contributions regardless of whether their income is tax-exempt. This is an administrative policy, and is implemented by submitting an application form {\it CPT124}, called {\em Application to Cover the Employment of an Indian in Canada under the Canada Pension Plan whose Income is Exempt under the Income Tax Act}. In other words, it seems that employers -- not employees --  are the ones who currently elect to have their First Nations employees contribute. But this is an administrative matter and is apparently not enshrined rigidly in the underlying legislation. Moreover, if a First Nations individual who is tax-exempt wished to contribute to the CPP, but their employer did not, the employee is required to make double the contributions {\em as if} they were self-employed. Why should they have to double the contribution rate? 

\vspace{0.1in}

We will shortly see that -- even with the idiosyncratic risk that would accompany abandoning the large national plan and instead participating in a relatively small but longevity-aligned plan -- individuals with a 10 year impairment in longevity could achieve undiminished utility at a reduced cost. But we emphasize that this conclusion rests on there being an alternative plan, to which employers would direct matching contributions. Without those matching contributions, it would take something like a 15 year longevity impairment before the alternative plan would become appealing.

\vspace{0.1in}

What are we left with then, beyond this administrative vagueness and complexity? It seems that some members of First Nations actually have the option to opt out of CPP -- but that administrative practice makes it harder if not impossible to do so. Thus, by changing or perhaps even tweaking administrative rules -- that is without any  legislative changes -- this group could leave the CPP, if they so desired. They could create their own retirement plan funded by both their and their employers' contributions, {\em at a much lower cost.}

\section{Modeling CPP vs. the Alternative}
\label{sec:model}

We will work with real (i.e. inflation adjusted) dollars. Since CPP payouts are indexed, and contributions are scaled up to account for inflation prior to retirement, this simplifies the modeling. The risk-free discount rate, based on the real (versus nominal) yield curve will be taken as constant and denoted $r$. Our model for CPP assumes a constant rate $c$ for (real) pensionable earnings, a contribution rate $\alpha$, and a replacement rate $\eta$, all in continuous time. So $\alpha c$ is what an individual contributes per year, and $\eta c$ is what they receive back per year in retirement. We assume that employers are obligated to match with a further $\alpha c$, so net contributions are $2\alpha c$.  We will work with power law (CRRA) utility $u(z)=\frac{z^{1-\gamma}}{1-\gamma}$ where $\gamma>0$, $\gamma\neq 1$ represents the risk aversion coefficient, allowing us to normalize from now on to $c=1$. For the case $\gamma=1$, see Appendix A. 

\vspace{0.1in}

To be very clear, the way we {\em think} about, model and capture the benefit of participating in the CPP risk sharing pool, is quite different from other research articles in the (Canadian) pension literature. First, we are not computing the Internal Rate of Return (IRR) of the DB pension, say for someone with average versus higher mortality, in the way that Groenewoud and Ponds (2025), as just one example, do. The expected IRR -- which tends to ignore the insurance value of pensions -- is likely to be lower than the risk free rate of interest. Groenewoud and Ponds (2025) report IRR values in the vicinity of 2\% to 3\% across a range of countries. We are not arguing that members of First Nations (who are tax-exempt) would get a better IRR {\em on average} by managing their own plan, which is unknown in advance. Rather, we focus on cost savings.

\vspace{0.1in}

Likewise, methodologically, we are not computing the ratio of benefits to contribution, as Schirle (2024) does, and arguing that the number is too low for members of First Nations. That approach effectively ignores the time value of money (TVM). In fact, since pension contributions are paid well before the benefits are ever received, it is quite likely that the total of all contributions (costs) incurred by a member of First Nations, even with higher mortality, will (still) be less than the benefits paid in the distant future. That lens, so to speak, will miss the problem we identify. Our methodology looks at the discounted expected utility of the entire pension benefit, adjusting for both when the benefit is received in time, and the probability of being alive to consume that benefit.

\vspace{0.1in}

Later on, in numerical work, we will take $\alpha=6\%$, and $r=1\%$ or $r=3\%$. The replacement rate $\eta$ will then be computed so as to be fair under our mortality assumptions for the general population, which will be Gompertz mortality with $b=10$ and $m=90$ (ie. a hazard rate of $\frac{1}{b}e^{\frac{x-m}{b}}$ at age $x$). In  Appendix A, this is shown to imply $\eta=0.3281$ when $r=1\%$, and $\eta=0.6514$ when $r=3\%$. To say this differently, we are modelling a stylized CPP in which there is no distinction between the pre-2019 rules, the post-2019 enhancements, the minutia of YMPE versus YAMPE, etc. In other words, we are also fully aware that up to the YMPE (\$71,300) the contribution rate is 5.95\%, and that between the YMPE and the YAMPE (\$81,200) the additional contribution is 4\% (in 2025), etc. We do, however, note that in our stylized CPP, a contribution rate of 6\% (very close to reality), which is matched by employers (reality), leads to a replacement rate of nearly one-third (the current plan for CPP), for population mortality rates, under a risk-free rate of 1\%, which is close to the current real rate in Canada. Ergo, we believe we have captured the essence of the CPP\footnote{In the list of other items we do not address (or even touch) in this paper, another one is this. Why does the valuation rate of $r=1\%$ fit the existing contribution (6\%) to payout (1/3) ratio rather well, yet the CPPIB fund is targeting a 4\% real rate and has earned much more than that over the last decade. In English, the CPP payout rate should be higher, given the rates they are earning in practice.}

\vspace{0.1in}

Longevity that is impaired, in contrast, relative to the base population, will be reflected by choosing lower values of the Gompertz parameter $m$. An impairment of $\delta$ years, for someone age $x$, means that their mortality hazard rate is that of someone age $x+\delta$ in the base population, and corresponds to using $m-\delta$ in place of $m$. So, while this is technically not precise given the nonlinearities in how mortality rates translate into life expectancies, we can consider a life expectancy gap of 10 years, meaning that $\delta=10$.

\vspace{0.1in}

In terms of working years and savings into the plan, we assume contributions for time $T$ (taken to be 40 in numerical work) starting at age $x_0$ (taken to $=25$ in numerical work) and running to age $x_1=x_0+T$.  Stated differently, workers are buying and investing in {\em deferred annuities}, a.k.a. ALDAs. To be conservative, and an assumption that is standard in all the money-worth ratio studies, we'll compute utilities assume that the individual has no other assets, savings, or income beyond the level of pensionable earnings. Extra assets or income would tend to make CPP even less appealing from a utility perspective. We are modelling people and participants who are solely dependent on the CPP for their financial survival. In that case then, the expected utility of lifetime retirement income, calculated at age $x_0$, is
\begin{equation*}
E\Big[\int_T^{T\lor\text{lifetime}}e^{-\rho t}u(\eta)\,dt\Big]=\int_{T}^\infty e^{-\rho t}{}_tp_{x_0} \frac{\eta^{1-\gamma}}{1-\gamma}\,dt=e^{-\rho T}{}_{T}p_{x_0}\frac{\eta ^{1-\gamma}}{1-\gamma}\bar a.
\end{equation*}
Here ${}_tp_x$ is the probability an individual of age $x$ will survive $t$ additional years, which for all of our models numerical results will be based on the Gompertz law of mortality, and where $\rho$ is a subjective discount factor, and $\bar a = \int_0^\infty e^{-\rho t}{}_tp_{(x_0+T)}\,dt = \int_0^\infty e^{-\rho t}{}_tp_{x_1}\,dt $. That then is the mathematics and utility of stylised CPP. So, what is the alternative? Well, an {\em equitable longevity risk sharing} plan could be designed in many ways, but we will analyse the following plan which could serve as a template for a dedicated Pension plan for First Nations. Participants, of which there are $n$ of them, all the same age and hazard rate, contribute at a rate $\alpha$ till age $x_1$ or death, whichever comes first. Notice that the number of participants and members -- which wasn't relevant in the guaranteed CPP -- is a critical aspect of the ELRiS. They are pooling among themselves, internally, in a self-supporting manner, but the larger the size of the pool (obviously), the lower the idiosyncratic risk. Either way, $n$ will be a critical variable. 

\vspace{0.1in}

The contributions of these $n$ members are matched by their employer, and the total is invested at the real interest rate $r$, which is the same exact rate used for valuing the CPP. We are assuming that the same opportunity set is available, the fees are the same and that no extra administrative charges are added. That is the true apples-to-apples comparison. 

\vspace{0.1in}

At age $x_1$, we denote the accumulated capital in the ELRiS fund, plan or account by the new variable $X$. Think of this as the fund value, consisting of contributions from employees and perhaps employers. These are the funds not invested in the CPP. At time $t=T+s$ (that is at age $x_1+s$) the fund then pays out at rate $d_sX$, and those payouts are shared equally among the $N_t$ survivors who are alive at that time. That then is the alternative design we are proposing for the fund, which in the limit as $n \rightarrow \infty$ would be identical in structure to the CPP if (and only if) mortality rates were identical. Of course, at this point astute readers will recognize that when $n << \infty$, the retirement income payments are non-constant, non-guaranteed and rather volatile, especially if $n$ is small. Moreover, this ELRiS design is a type of tontine fund or arrangement, so this paper doesn't propose a new or novel design, but a modification.\footnote{This is a product which has also been analyzed and investigated by Bernhardt \& Donnelly (2019), Chen et al (2019, 2020, 2022, 2024), Dagpunar (2021), Forman \& Sabin (2015), Forsyth et al (2024), Gemmo et. al. (2020), Hieber \& Lucas (2022), Pflaumer (2022), Sabin (2010), Stamos (2008), Weinert \& Gr\"undel (2020), as well as Winter \& Planchet (2022). They also fall under the title of Pooled Annuity Funds, see for example Donnelly (2015) or Piggott et al (2005), Qiao \& Sherris (2013). As with the {\em annuity puzzle} the literature is large and growing, and here we simply acknowledge that our proposal to create a First Nations Pension Plan could be structured in many ways, such as those suggested in the various papers cited. The key is to pool like-with-like.} Again, our objective is to compute how much members of First Nations (with impaired mortality) would have to contribute to the fund or account to create an income stream with the same level of discounted expected utility. 

\vspace{0.1in}

In the ELRiS plan, product design -- that is the payout rates to survivors -- amounts to choosing $d_s$, subject to the natural budget constraint $\int_0^\infty e^{-rs}d_s\,ds=1$. This $d_n$ will also be {\em actuarially fair} as the size of the pool $n \rightarrow \infty$. As noted, we choose $d_s$ to maximize the expected lifetime utility of a reference individual. Let $X=2\alpha Y$, recalling the doubling of contributions by the employer. In other words, $Y$ is the future value of the individuals portion of the fund, and $2Y$ after the employer match is the future value of the entire fund. This must be symmetric to the CPP. Also, the new variable $E_s$ denotes an expectation conditional on the reference individual being alive at time $t=s+T$, so $N_{s+T}-1$, which  is conditionally Binomial with mean $(n-1){}_{s+T}p_{x_0}$. The expected discounted lifetime utility ({\bf U}) for the reference individual under this ELRiS is then
\begin{equation*}
{\bf U} =\int_0^\infty \frac{d_s^{1-\gamma}}{1-\gamma}e^{-\rho(s+T)}{}_{s+T}p_{x_0}E_s\Big[\big(\frac{X}{N_{s+T}}\big)^{1-\gamma}\Big]\,ds
=e^{-\rho T}{}_Tp_{x_0}\frac{(2\alpha)^{1-\gamma}}{1-\gamma}\int_0^\infty d_s^{1-\gamma}e^{-\rho s}\beta_s\,ds,
\end{equation*}
where $\beta_s={}_{s}p_{x_1}E_s\Big[\big(\frac{Y}{N_{s+T}}\big)^{1-\gamma}\Big]$. As far as notation, $s$ is just an index variable and the utility is being discounted to time zero. 

\vspace{0.1in}

Lagrange multipliers now imply that the optimal $d_s=\lambda [e^{(r-\rho)s}\beta_s]^{\frac{1}{\gamma}}$ for some $\lambda$. We will simplify this by taking $\rho=r$, so by the budget constraint, $\lambda=1/\int_0^\infty e^{-\rho s}\beta_s^{\frac{1}{\gamma}}\,ds$. And, as in Milevsky-Salisbury (2015), which first introduced this type of tontine into the literature, the above utility then becomes
\begin{equation*}
{\bf U} =e^{-\rho T}{}_Tp_{x_0}\frac{(2\alpha)^{1-\gamma}}{1-\gamma}\Big(\int_0^\infty e^{-\rho s}\beta_s^{\frac{1}{\gamma}}\,ds\Big)^\gamma.
\end{equation*}

\vspace{0.1in}

Let $\bar\alpha$, an alpha not to be confused with the annuity factor $a$, be a contribution rate for which the CPP alternative yields the same expected utility as CPP itself. Though we will focus on comparing $\alpha$ with $\bar\alpha$ as our measure of inequity, we could also have considered the recovery $\bar\eta$ at which the alternative tontine, using contribution rate $\alpha$, would provide the same expected utility as CPP. In the limit, of course, as the number of participants $n \rightarrow \infty$ the structure would be identical to the guaranteed annuity from the CPP.
In either case, we obtain the relationship
\begin{equation*}
\Big(\frac{2\bar\alpha}{\eta}\Big)^{1-\gamma}
=\frac{\bar a}{\Big(\int_0^\infty e^{-\rho s}\beta_s^{\frac{1}{\gamma}}\,ds\Big)^\gamma}
=\Big(\frac{2\alpha}{\bar \eta}\Big)^{1-\gamma}.
\end{equation*}
Thus $\bar\alpha$ will be easy to calculate, once we compute the quantities $\beta_s$. And of course, $\bar\eta=\eta\frac{\alpha}{\bar\alpha}$.

\vspace{0.1in}

From a purely technical or mathematical perspective, comparing this with the algorithm described in Milevsky-Salisbury (2015), we see that this presents a harder computational problem than with the latter, since that paper had a decumulation phase only. Whereas now we must deal with both accumulation and decumulation, complicated by the fact that $X$ and $N_{s+T}$ are dependent quantities. Since this paper is primarily focused on the financial economics (and policy aspects) of the ELRiS, we address this technical issue in Appendix A.

\section{Numerical Results \& Intuition}
\label{sec:results}

Table \#\ref{tab:mytable}  contains the numerical results of our analysis. Although there are a number of free parameters that we could vary and investigate its impact on the contribution rate, we focus on two numbers of specific interest. The first is the risk aversion level $\gamma$, which drives the extent to which the insurance element of the pension is appreciated, and $n$, which is the size of the pool. The output, denoted by $\bar{\alpha}$ -- once again -- is the percent of wages that an impaired life ($\bar{m} \leq m$) could contribute to an {\em equitable longevity risk sharing} (ELRiS) pool for the Indigenous, and still obtain the same level of discounted lifetime utility. We used two different interest rates, one being the (more reasonable, current) $r=1\%$ and the second being a much higher $r=3\%$, which, recall, is stated in real inflation-adjusted terms. 

\vspace{0.1in}

When the risk-free rate in the system is changed, then obviously the `fair' CPP replacement rate also changes, and that is displayed as well. Notice that under an $r=3\%$ system-wide valuation rate, the retirement replacement rate (for a 6\% contribution to CPP, matched by employers) is 65.14\%. For $r=1\%$, the basis of most of our displayed and discussed results, the replacement rate is 32.8\%. The utility-equivalent {\em replacement rate} is denoted by $\bar{\eta}$, and is the fraction of (average) lifetime wages that should be paid out in the {\em equitable longevity risk sharing} (ELRiS) fund, assuming the impaired life contributed at (the same) rate $\alpha.$  So, the parameter $\bar{\alpha}$ is expressed in the language or units of contribution rates, and the $\bar{\eta}$ in terms of increased benefits. To put this all in relative perspective and circle back, our stylized Canadian Pension Plan (CPP) would be fair under an $\alpha=6\%$ contribution rate and $\eta$=32.8\% replacement rate, {\em if} contributions were made over a period of $T=40$ years, under a fixed real interest rate of $r=$1\%. And, provided we assume that the population $m$ (modal value of life) is 90 years (ie for the base Canadian population), and the pension plan is priced using that basis. 

\vspace{0.1in}
\begin{center}
{\bf [Table \#\ref{tab:mytable} goes here]}
\end{center}
\vspace{0.1in}

Our choice of baseline preference parameters were directed by the recent empirical work of Laibson et al (20024). Here is how to interpret the results from one small section of the table. Let's begin with a mid-point coefficient of relative risk aversion $\gamma = 2$, with an impaired life of $\bar{m} = 80$ years, which is a ($\delta=10$) decade lower than the underlying population basis used for pricing. With the duality of contribution versus replacement rates in mind, we can read and interpret the results in two ways. First, the impaired life could attain equivalent lifetime utility by contributing two-thirds, that is $\bar{\alpha}=$ 3.82\%, to an ELRiS consisting of $n = 30$ members. The (expected) replacement rate in the ELRiS fund would be the $\eta=32.8\%$ of the CPP, and the (expected) utility would be the same as if they had contributed the $\alpha=6\%$ to the guaranteed pension plan.  The second interpretation is that, based on the identity $\bar\eta=\frac{\alpha\eta}{\bar\alpha}$, the CPP's contribution rate of $\alpha=6\%$, if made instead to the ELRiS, should translate to a replacement rate of $\bar\eta=51.6\%$. This is, of course, significantly higher than $\eta$. 

\vspace{0.1in}

To be very clear, the actual payout (consumption) profile in the {\em equitable longevity risk sharing} fund will fluctuate based on the (idiosyncratic risk, and) actual number of survivors. Using the language of Boon,  Bri\`ere and Werker (2020), we are arguing this risk should be 'borne' by the pool. Bearing this risk will be especially severe or extreme in case when $n$, the original size of the pool at the time the plan is set up, is relatively small. As noted in the modelling section, our alternative to the CPP is a type of pooled fund (i.e tontine) in which payments to survivors are (highly) unpredictable at (very) advanced ages. Nevertheless, the expected discounted utility -- that is 3.82\% to the fund versus 6\% to the CPP -- would be exactly the same. Contribute less money, but earn the same utility. In the other direction, a contribution rate of 6\% to the CPP for someone with impaired longevity should lead to an expected (remember, nothing is guaranteed) replacement rate that is higher than the $\eta=$ 32.8\% that is guaranteed. This latter perspective -- First Nations members should be getting a better replacement rate -- offers a different viewpoint on the {\em welfare loss} from participating in the CPP versus a self-supporting plan for First Nations.

\vspace{0.1in}

In the same section of the table, that is for $n=30$, under a longevity gap of $m - \bar{m} = 20$ years, the utility-equivalent contribution rate falls to: $\bar{\alpha}=$1.78\%. This is low enough that even in the absence of employer matching, the ELRiS is favourable (ie even if the individual needed to double their own contributions). Equivalently, the utility-equivalent replacement rate (assuming employer matching) rises to $\bar\eta=110.6\%$ which actually exceeds the 100\% rate of labour income.

\vspace{0.1in}

We should note that increasing the coefficient of risk aversion from $\gamma=2$ to $\gamma=10$, which is at the upper edge of acceptable risk aversion parameters, the equivalent contribution rate moves only slightly to $\bar{\alpha}=$ 2.25\% (or $\bar\eta=87.3\%$). The interpretation here is that even a (very) longevity risk-averse individual with an impaired life of $\bar{m}=70$ years, would be content with a contribution rate of approximately 2.25\% versus 1.78\%, under moderate risk aversion. In fact, on a qualitative level, the results aren't dramatically sensitive to the coefficient of risk aversion or the risk-free rate, although the equivalent replacement rate ($\bar{\eta}$) is much higher when the interest rate is increased. The intuition is that (larger) discounting allows for much greater benefit replacement rates in both the CPP and the ELRiS for members of groups with impaired longevity, such as First Nations.

\vspace{0.1in}

A byproduct of our analysis, which appears under the title of systematic withdrawal plan (SWIP) in the table, is that if longevity is sufficiently impaired ($\bar{m} << m$), the individual is better off not participating in the CPP at all, even if it means doing it alone. To be very clear, since all contributions are matched by the employer (in the CPP), this assumes the employer will direct the matching contributions to the ELRiS. For example, under the same level of moderate risk aversion $\gamma=2$, under the $n=1$ panel, with $\bar{m}=70$, the utility equivalent contribution rate $\bar{\alpha}$ = 5.42\%, which is less than the CPP contribution rate of $\alpha=$ 6\% (and so $\bar\eta=36.3\%$ is higher than the CPP's $\eta=32.8\%$). This also means that the individual will optimize their utility by consuming less as they age, in direct proportion to their survival rate (squared), but still be better off than had they participated in the fund. We are not advocating self-annuitization or solo-decumulation as an alternative to participating in the CPP. Rather, this is meant to show the limited benefits of {\em equitable longevity risk sharing} when the individual is effectively subsidizing those with better longevity prospects. In fact, one could write down an expression for the longevity gap $m-\bar{m}$, for $n=1$ and a given $r,\gamma$, that leads to $\bar{\alpha}<\alpha$.

\begin{figure}
\vspace{-0.1in}
\caption{Simulated Pension Payouts}
\label{fig2}
\vspace{0.5in}
\begin{center}
\includegraphics[width=0.45\textwidth]{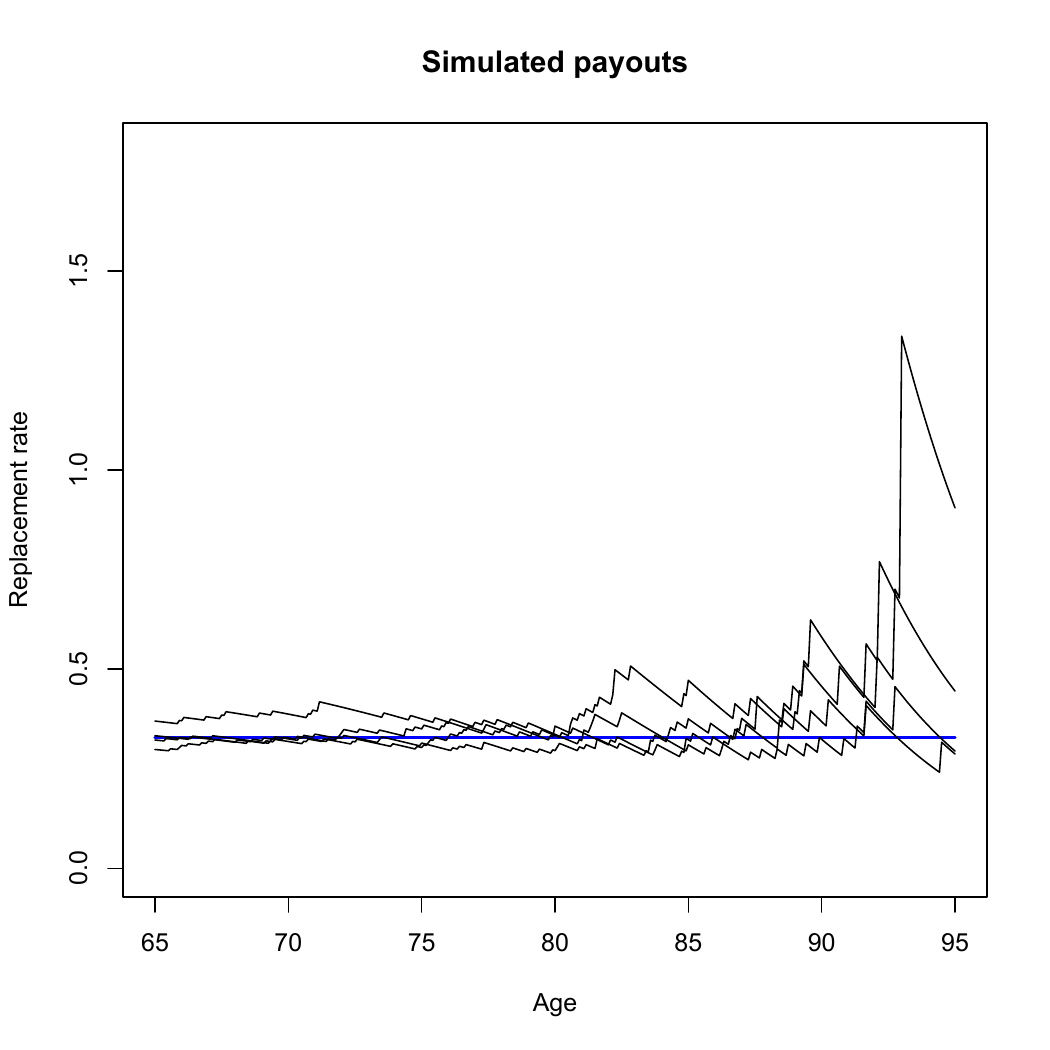} 
\includegraphics[width=0.45\textwidth]{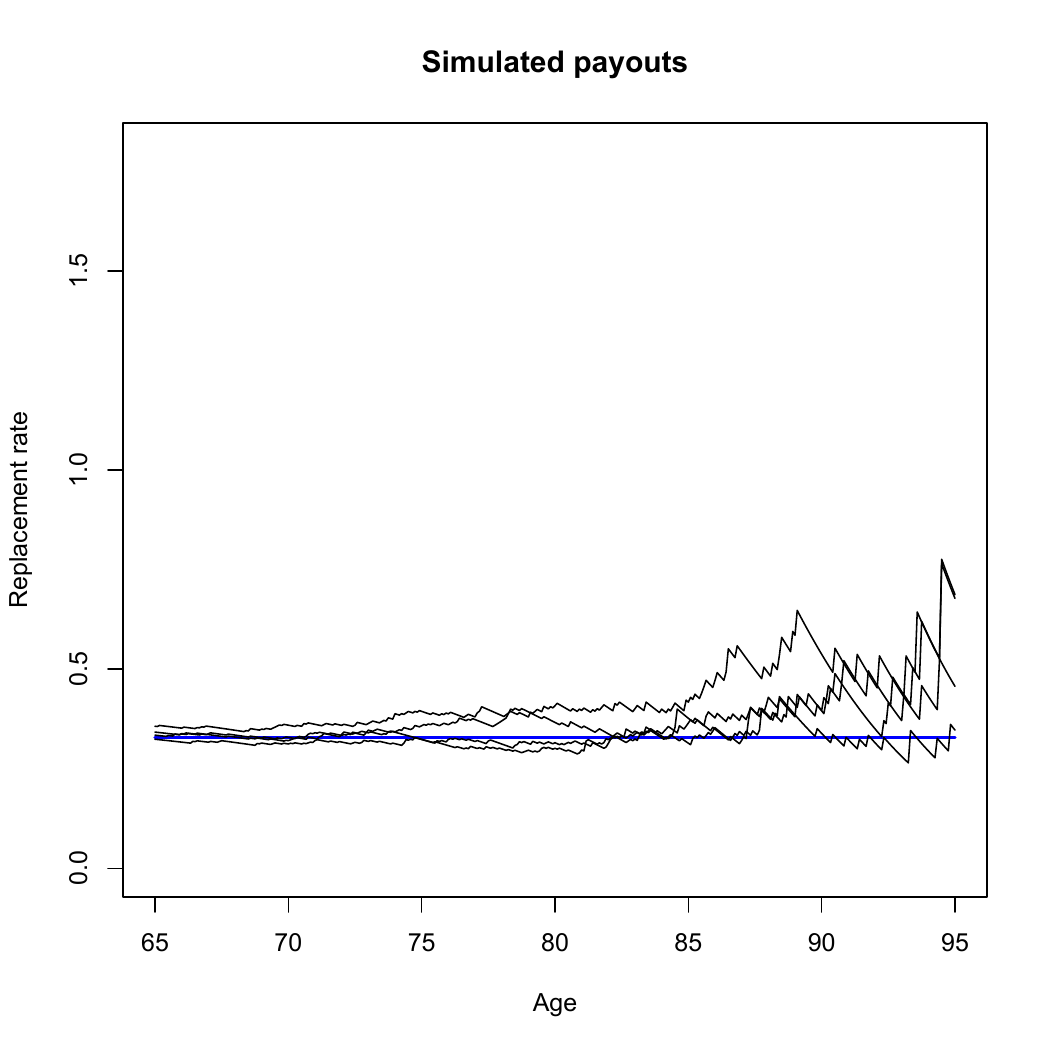} 
\end{center}
\end{figure}

\vspace{0.1in}
\begin{center}
{\bf [Figure \#\ref{fig2} goes here]}
\end{center}
\vspace{0.1in}

A few items remain for a proper understanding of how the ELRiS fund or account might operate as an alternative to the CPP. One is gaining an understanding of the sample paths of consumption. To that end, Figure \#\ref{fig2} displays four simulated paths of yearly income, normalized to a pensionable income of $c=1$, for an ELRiS with $n=30$, and then for one with $n=150$. The plan is optimized assuming a high risk aversion of $\gamma=10$, and uses the calculated contribution rate $\bar\alpha=0.0433$, which is from Table \#\ref{tab:mytable}. The horizontal line represents the CPP replacement rate of $\eta=0.3281$ corresponding to $r=1\%$ and $\alpha=0.06$. We assume a 10-year impairment of mortality, i.e., $\bar m=80$. Even with $n=30$ (left) and $n=150$ (right), the figure shows reasonably stable income through age 85, mostly exceeding $\eta$. Wilder oscillations after that point are due to the high probability of a small number of survivors, as in any tontine fund arrangement. Alternate designs could be considered that might damp down these oscillations to some extent. With even larger $n$, payments would be even more stable. We see an upward jump with each death, and a decline between deaths. The decline is driven by the numerator ($d_x$) at the core of the cash-flow available for distribution. The variability becomes largest between ages 85 and 95, declining after that point as the scenario of a single survivor becomes dominant.

\vspace{0.1in}

\begin{figure}
\vspace{-0.1in}
\caption{Percentiles (20, 40, 60, 80) of Replacement Rates}
\label{fig3}
\vspace{0.5in}
\begin{center}
\includegraphics[width=0.45\textwidth]{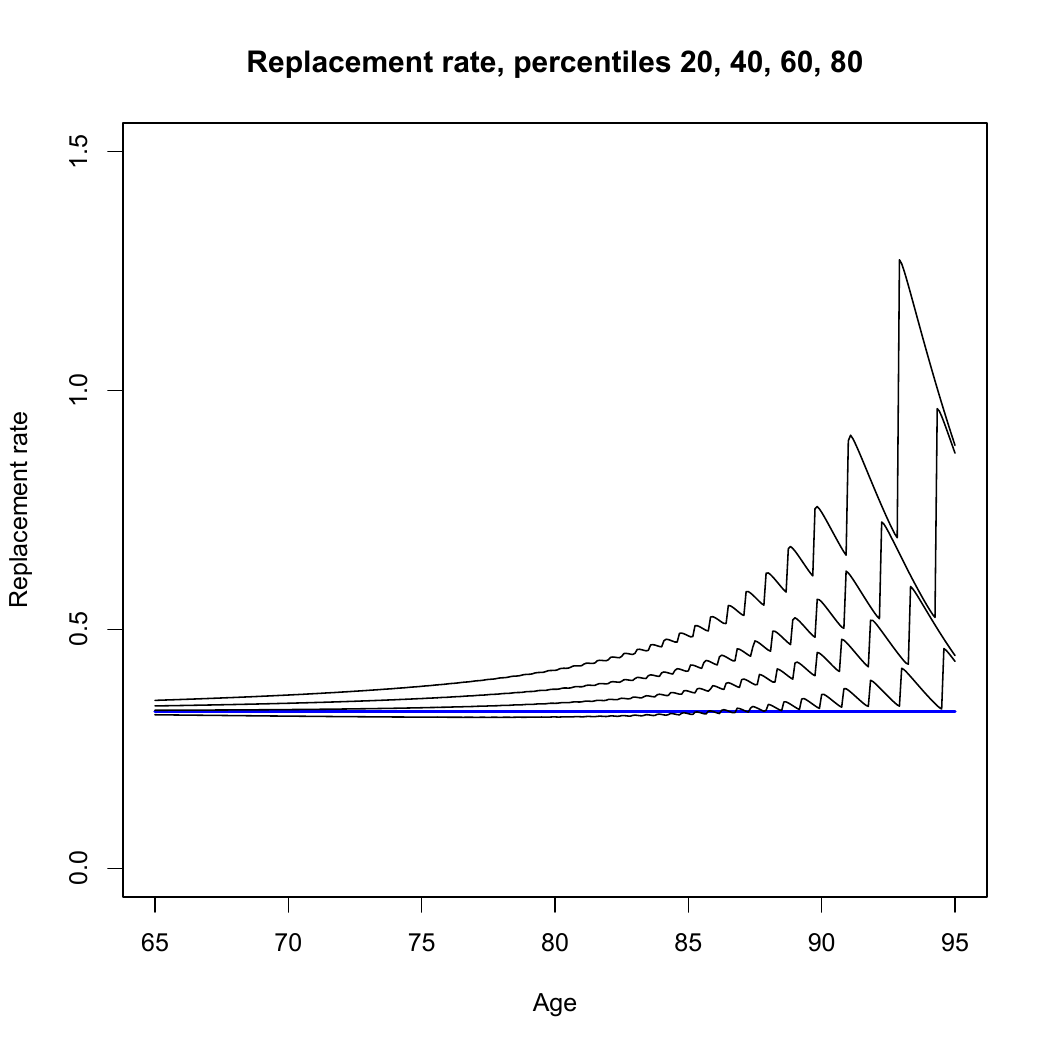} 
\includegraphics[width=0.45\textwidth]{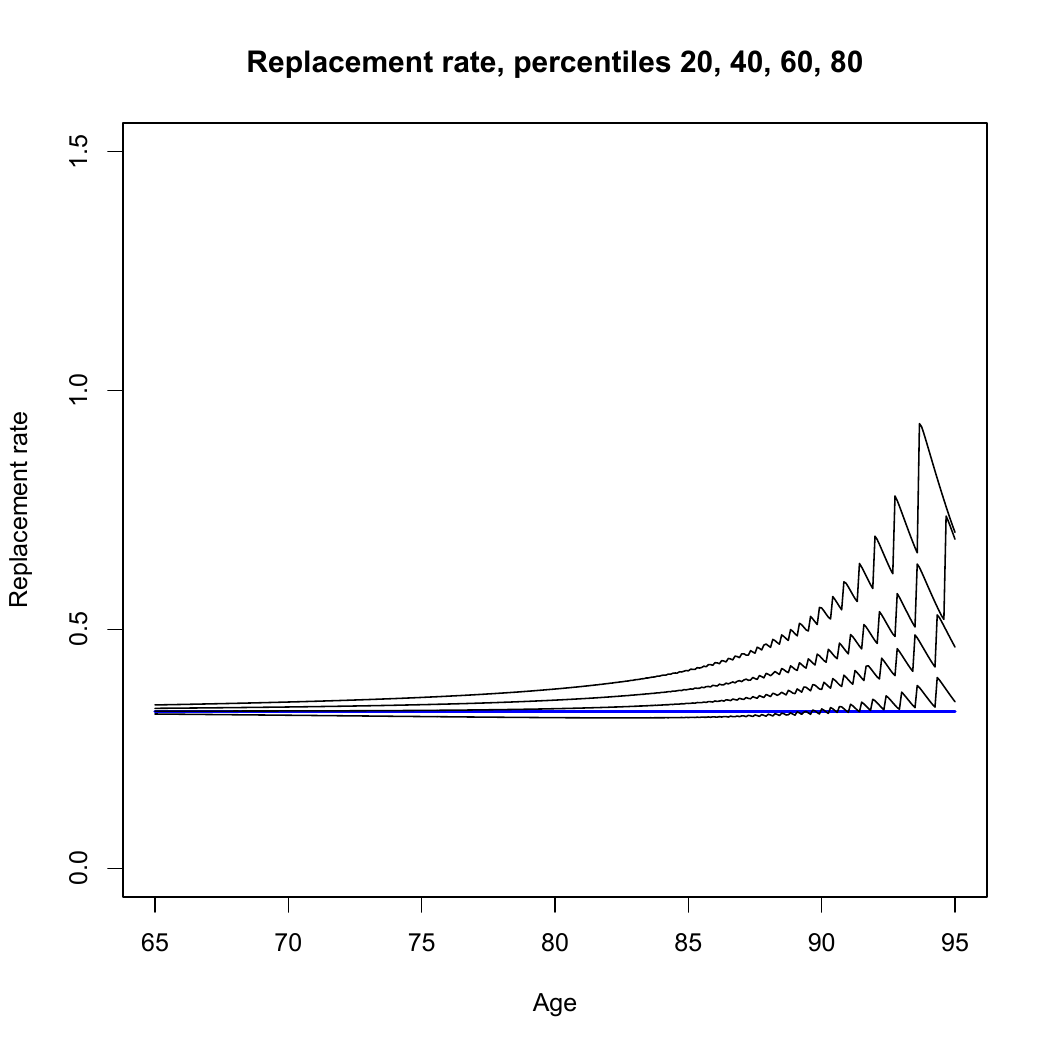} 
\end{center}
\end{figure}

\begin{center}
{\bf [Figure \#\ref{fig3} goes here]}
\end{center}

\vspace{0.1in}

Continuing towards a better understanding of the income or consumption paths, Figure \#\ref{fig3} shows the 20'th, 40'th, 60'th and 80'th percentiles of yearly income, normalized to a pensionable income of $c=1$, for a pool with $n=30$, and then with $n=150$. The plan is optimized assuming a high risk aversion of $\gamma=10$, and uses the calculated contribution rate $\bar\alpha=0.0433$; The horizontal line represents the CPP replacement rate of $\eta=0.3281$ corresponding to $r=1\%$ and $\alpha=0.06$. We assume a 10 year impairment of mortality, ie. $\bar m=80$. Even with $n=30$ (left), the figure shows reasonably stable income through age 85, mostly exceeding $\eta$. Oscillations after that point are due to the high probability of a small number of survivors. Once again, alternate designs to the one we have proposed could be considered, and those might damp down these oscillations. Remember, once again, with larger $n$, payments would be even more stable and the $(n=150)$ figure on the right indicates how the fluctuations are reduced with larger pools.  Be it $30$ or even $150$, these are small pools, and the size of the First Nation population is in the hundreds of thousands, but this tells us that these plans can be a reasonable alternative even with a very small number of participants. Once again, we see an upward move each age at which the dominant number of survivors in a percentile drops by 1, and a decrease in payment between such upticks. The variability becomes largest between ages 85 and 95, declining after that point as the scenario of a single survivor becomes dominant.

\vspace{0.1in}
\begin{center}
{\bf [Figure \#\ref{fig4} goes here]}
\end{center}
\vspace{0.1in}

Figure \#\ref{fig4} plots the contribution rate $\bar\alpha$ against the impaired value of $\bar m$, assuming $r=1\%$, and optimized for $\gamma=2$. We show various pool sizes $n$. This particular figure is meant to answer the question: how does the interaction between the size of the impairment ($\delta$ ) and the size of the pool ($n$) affect the contribution rates that would lead to the same level of expected utility? This figure also answers the question: What if members of the ELRiS have lower morality and better longevity prospects? This obviously doesn't reflect reality, but does aid in developing an intuition for contribution rates that generate the same level of utility as a guaranteed pension plan. The vertical line represents the $m=90$ of the base population, and the horizontal line shows the assumed CPP contribution rate of $\alpha=6\%$. Of course, these cross the $n=\infty$ curve at the same point. After all, if ELRiS members have the same exact longevity prospects -- and the size of the pool goes to infinity -- the contribution rate should be the same 6\%. Here we also show a full range of $\bar m$, but omit $n=30$ for clarity.

\vspace{0.1in}
\begin{center}
{\bf [Figure \#\ref{fig5} goes here]}
\end{center}
\vspace{0.1in}

Figure \#\ref{fig5} is identical to Figure \#\ref{fig4}, except that we zoom in on the $\bar m$ of interest. And that at this more granular level, we are able to include the case $n=30$. The main intuitive take-away from this figure is to show (once again) and in yet another manner that individuals with impaired mortality, that is with longevity prospects that are 5, 10, 15, 20, years lower than the baseline population on which the plan is based, can in fact contribute {\em less} than 6\% to the alternative plan, and still obtain the exact same level of utility. 

\vspace{0.1in}

Finally, earlier, we discussed a very common (though we believe, flawed) way of thinking about the generosity of a national pension plan, based on expected benefits relative to contributions. Appendix B to this paper contains some analytic insights into the ratio of expected (un-discounted) benefits $E[B]$, to expected (un-discounted) contributions $E[C]$, which we define as the generosity $G=E[B]/E[C]$ for the stylized CPP. In particular, some numerical values of this generosity metric are as follows. When the individual mortality mode ($m=90$) is equal to the population mode in our model, and the CPP is actuarially fair, the generosity ratio is $G=2.79$, in line with what is reported in the (Canadian) literature. For example, see the tables in Schirle (2024, pg. 665). However, using the same generosity metric under a modal value of 70 (or 80) years, versus the population 90 years, the generosity of the CPP plummets to $G=0.70$ (or $G=1.64$), respectively. And, while this undiscounted metric is imperfect to a financial economist, we should note that under our {\em equitable longevity risk sharing} (ELRiS) plan, the generosity numbers – under the same $r=1\%$ interest rate and $\alpha=6\%$ contribution rate, under a risk-aversion parameter $\gamma=10$ — increase to $G=2.45$, for $m=70$ modal value of life. This, of course, is higher compared to our stylized CPP.  So, an ELRiS vehicle or structure is not only {\em equitable} using proper utility-based and actuarial funding metrics, it is also more generous.

\section{Conclusion}
\label{sec:conclusion}

This paper is long as is, so we refrain from repeating ourselves and conclude as follows. At the core, we argue that identifiable groups with substantially elevated mortality rates relative to the rest of the population are better off creating their own (yes, smaller) but {\em equitable longevity risk sharing} pools for retirement income. Our case study -- or perhaps a trial balloon for this idea -- is the First Nation peoples of Canada, some of whom already have the option to do so. In other words, they might consider seceding from the guaranteed national pensions in favour of smaller, demographically aligned risk pools. Our arguments are utility-maximizing in nature rather than political or sociological. They are solely traced to the large gap in mortality between the Indigenous peoples of Canada and the non-Indigenous. If other arguments -- beyond longevity impairments -- can be marshalled to allow them to create, contribute and manage their own pension plan, the case for an ELRiS is even stronger.  For example, under CPP (legislation) spouses continue to receive reduced CPP benefits -- though it is something we didn't model in the paper -- and children may receive benefits under limited circumstances. But, the life annuity benefits end when the insured and spouse have both died. That is how longevity risk pooling works. But, First Nations often have a much broader conception of family, which includes not only parents and grandparents, children and grandchildren, but even distant nieces, cousins and nephews. A self-administered FN plan would enable Indigenous members with a broader conception of family within society to continue (albeit lower) benefits for longer, paid not only to spouses but other dependents as well.

\vspace{0.1in}

We don't want to overstate our case, but the motivation purely on mortality considerations alone is quite strong.  Under typical utility preferences, a member of First Nations with a 10-year life expectancy gap, relative to the Canadian population, could attain equivalent lifetime utility by contributing a mere {\em two-thirds} to such a plan even if they are pooled with 30 members. For a longevity gap of 20 years, such as between an Indigenous male versus a non-Indigenous female, the required contribution rate falls to less than {\em a third.} The difference between the nearly 6\% statutory contribution rates to CPP and rates needed within our plan captures the implicit subsidy from Indigenous (males) to non-Indigenous (females), which is obviously unfair and inequitable. In sum, we hope this paper jump-starts a conversation to effect a required change in the status quo.

\newpage 

\section{Appendix A}
\label{sec:appendix}

\begingroup
\singlespacing

We start by calibrating $\eta$ to be fair for the base population. This will happen if the present value of CPP contributions matches the present value of CPP benefits. In other words (recalling that we are including employer matching contributions) if
\begin{equation*}
2\alpha \int_0^T c e^{-r t}{}_tp_{x_0}\,dt = \eta \int_T^\infty ce^{-r t}{}_tp_{x_0}\,dt.
\end{equation*}
With $\alpha=6\%$ and $\rho=1\%$ this implies $\eta=0.3281$, whereas $\rho=3\%$ yields $\eta=0.6514$.

\subsection{Finding $\beta_s$}
Our main problem is to compute $\beta_s$. When $n=1$, this is simple as the only states contributing to the utility are those for which $Y=\int_0^T e^{\rho(T-t)}\,dt=\frac{e^{\rho T}-1}{\rho}$.

In general, 
\begin{equation*}
Y=\int_0^TN_te^{\rho(T-t)}\,dt.
\end{equation*}
We also know that $N_t$ and $N_{s+T}$ are conditionally independent, given $N_T$. Therefore 
\begin{equation*}
\beta_s={}_{s}p_{x_1}
\sum_{\ell=1}^n E_s\Big[\frac{1}{N_{s+T}^{1-\gamma}}\mid N_T=\ell\Big]E_s\Big[\big(\int_0^T N_te^{\rho(T-t)}\,dt\big)^{1-\gamma}, N_T=\ell\Big].
\end{equation*}
Let $\beta_{\ell,s}$ and $b_\ell$ respectively denote the first and second expectations inside the sum. Note that  $b_\ell$ does not actually depend on $s$. The $\beta_{\ell,s}$ term is easy to compute,  as
\begin{equation*}
\beta_{\ell,s}=\sum_{k=0}^{\ell-1}\Big(\frac{1}{1+k}\Big)^{1-\gamma}\binom{\ell-1}{k}{}_sp_{x_1}^k(1-{}_sp_{x_1})^{\ell-1-k}.
\end{equation*}

Our computation of the $b_\ell$ used a Markov chain approximation to the dynamics of $N_t$ and
\begin{equation*}
dY_t=\rho Y_t\,dt + N_t\,dt,
\end{equation*}
(or equivalently, $Y_t=\int_0^t N_se^{\rho (t-s)}\,ds$). We discretize the values for $t$ and $Y_t$ and calibrate the transition rates from $(t,y,k)$ to one of $(t+\delta, y,k)$, $(t+\delta,y,k-1)$, $(t+\delta, y+\Delta,k)$, or $(t+\delta,y+\Delta,k-1)$ to achieve the correct mean values. And then compute $b_\ell$ via the Markov chain's backwards equation. This ends up being the most time-consuming part of the computation of the $\bar\alpha$, each of which took between 2 and 3 server hours for $n=30$ at a resolution of 8,000 $t$'s and 4,000 $y$'s. In fact, we ran also ran at lower resolutions, and then applied Richardson extrapolation to yield the reported numbers.

\subsection{The case $\gamma=1$}
When $\gamma=1$ we use logarithmic utility instead. CPP now has expected utility
\begin{equation*}
\int_T^\infty e^{-\rho t}{}_tp_{x_0} \log(\eta)\,dt = e^{-\rho T}{}_Tp_{x_0}\bar a \log(\eta).
\end{equation*}
Our ELRiS has expected utility
\begin{multline*}
{\bf U} = \int_0^\infty e^{-\rho(s+T)}{}_{s+T}p_{x_0}E_s\Big[\log\big(\frac{d_s X}{N_{s+T}}\big)\Big]\,ds \\
 = e^{-\rho T}{}_Tp_{x_0}\Big[\bar a\log(2\alpha)+\int_0^\infty e^{-\rho s}\beta_s\,ds + \int_0^\infty  e^{-\rho s}{}_sp_{x_1}\log(d_s)\,ds\Big],
\end{multline*}
where 
$\beta_s={}_{s}p_{x_1}E_s\Big[\log \frac{Y}{N_{s+T}}\Big]$.
The same budget constraint on $d_s$ as before leads to $d_s=\frac{1}{\bar a} {}_sp_{x_1}$ and expected utility
\begin{equation*}
{\bf U} = e^{-\rho T}{}_Tp_{x_0}\Big[\bar a\log\big(\frac{2\alpha}{\bar a}\big)+\int_0^\infty e^{-\rho s}\beta_s\,ds + \int_0^\infty  e^{-\rho s}{}_sp_{x_1}\log({}_sp_{x_1})\,ds\Big].
\end{equation*}
As before, this gives rise to alternative rates $\bar\alpha$ and $\bar\eta$ determined by 
\begin{align*}
\log(\frac{2\bar\alpha}{\eta})&=\log(\frac{2\alpha}{\bar\eta})=\log\bar a -\frac{1}{\bar a}\Big[\int_0^\infty e^{-\rho s}\beta_s\,ds + \int_0^\infty  e^{-\rho s}{}_sp_{x_1}\log({}_sp_{x_1})\,ds\Big]\\
&=\log\bar a-E[\log(Y)] +\frac{1}{\bar a}\Big[\int_0^\infty e^{-\rho s}{}_sp_{x_1}E_s[\log N_{s+T}]\,ds - \int_0^\infty  e^{-\rho s}{}_sp_{x_1}\log({}_sp_{x_1})\,ds\Big]
\end{align*}
We find $\beta_s$, using a similar scheme to that of the preceding case. Though it simplifies, since the log decouples $X_0$ from $N_{s+T}$, and instead of finding $b_1,\dots,b_n$, we only require $E[\log(Y)]$. 

\subsection{Limiting case $n=\infty$}
As $n\to\infty$, the accumulated wealth per initial investor $=\frac{X}{n}\to 2\alpha \int_0^T{}_tp_{x_0}e^{\rho(T-t)}\,dt$ which we denote $2\alpha y$. Individuals receive $\frac{d_sX}{N_{s+T}}\to \frac{2\alpha y}{{}_{s+T}p_{x_0}}d_s$ so we want to maximize expected utility
\begin{equation*}
{\bf U} = \int_0^\infty e^{-\rho(s+T)}{}_{s+T}p_{x_0}\frac{1}{1-\gamma}\Big(\frac{2\alpha y}{{}_{s+T}p_{x_0}}d_s\Big)^{1-\gamma}\,ds.
\end{equation*}
This leads to $d_s=\lambda \cdot{}_sp_{x_1}$ and the budget constraint implies $\lambda=\frac{1}{\bar a}$. So we get an expected utility of

\begin{align*}
{\bf U} = e^{-\rho T}{}_{T}p_{x_0}\frac{(2\alpha)^{1-\gamma}}{1-\gamma}\int_0^\infty e^{-\rho s}{}_sp_{x_1}\Big(\frac{y}{{}_Tp_{x_0}\cdot\bar a}\Big)^{1-\gamma}\,ds \\
= e^{-\rho T}{}_{T}p_{x_0}\frac{(2\alpha)^{1-\gamma}}{1-\gamma}\bar a\Big(\frac{y}{{}_Tp_{x_0}\cdot\bar a}\Big)^{1-\gamma}.
\end{align*}

Therefore 
\begin{equation*}
\frac{2\bar \alpha}{\eta}=\frac{2\alpha}{\bar\eta}=\frac{\bar a\cdot {}_Tp_{x_0}}{y}=\frac{\bar a e^{-\rho T}\cdot {}_Tp_{x_0}}{\int_0^Te^{-\rho t}{}_tp_{x_0}\,dt}.
\end{equation*}
Which, of course, no longer depends on $\gamma$. We also refer readers to Section 6.2 of Chen and Rach (2022), where similar computational issues are addressed.
\newpage

\section{Appendix B}
\label{sec:appendixb}

As noted in the body of the paper, the pension economics literature often uses a simple ratio of benefits to contributions to measure the generosity of a pension system to various sub-groups within the population, or across populations and countries. This metric ignores the time value of money, and thus exaggerates the magnitude of benefits which occur in the distant future, relative to the immediate costs. Nonetheless, we compute and report those values for the ELRiS plan and compare with the appropriate estimates for the CPP, under impaired mortality, to show that even using this (flawed) metric, ELRiS is more generous. 

\vspace{0.1in}

To be precise, the quantity we (are defining as Generosity and) are interested in is:
$$
G:=\frac{E[\text{total un-discounted benefits}]}{E[\text{total un-discounted contributions}]}.
$$
Recall that in our (simple) model, everyone earns the maximum (capped) salary and is thus entitled to the maximum benefit, so the salary level $c$ scales out of the ratio for $G$, and we will take $c=1$

\vspace{0.1in}

For CPP,  total contributions take place at a rate $\alpha$, so the denominator of $G$ is
$$
E\Big[\int_0^T \alpha 1_{\{\text{alive at time $t$}\}}\,dt\Big]=\alpha\int_0^T{}_tp_{x_0}\,dt.
$$
The numerator becomes
$$
E\Big[\int_T^\infty \eta 1_{\{\text{alive at time $t$}\}}\,dt\Big]=\eta\int_T^\infty{}_tp_{x_0}\,dt.
$$
Note that we are computing an age $x_0$ life expectancy in retirement, which will be lower than the corresponding age $x_1$ life expectancy value. Either way, with our choice of parameters, and with fully impaired $m=70$ (resp. 80 and 90 years) longevity, these imply ratios of 0.70 (resp. 1.64 and 2.79)
In other words, the generosity of the CPP in our theoretical model -- for the segment of the population with complete longevity -- is 2.79, which can be compared to the values between 3 and 4 reported by Schirle (2024, pg. 664), albeit with a slightly different methodology. 

\vspace{0.1in}

Turning to the {\em equitable longevity risk sharing} (ELRiS) design we are proposing for those with impaired mortality, the expected contributions are $\bar\alpha\int_0^T{}_tp_{x_0}\,dt$. The expected benefit is computed as before for utility, namely, it
$$
=E\Big[\int_0^\infty \frac{d_sX}{N_{s+T}}1_{\{\text{alive at time $s+T$}\}}\,ds\Big]
=2\bar \alpha{}_Tp_{x_0}\int_0^\infty d_s\hat\beta_s\,ds
=\frac{2\bar\alpha{}_Tp_{x_0}}{\int_0^\infty e^{-rs}\beta_s^{\frac{1}{\gamma}}\,ds}\int_0^\infty \beta_s^{\frac{1}{\gamma}}\hat\beta_s\,ds
$$
where $\hat\beta_s={}_sp_{x_1}E_s\Big[\frac{Y}{N_{s+T}}\Big]$.

\vspace{0.1in}

As before,  $\hat\beta_s={}_sp_{x_1}\sum_{\ell=1}^n \hat\beta_{\ell,s}\cdot \hat b_\ell$, where
$$
\hat \beta_{\ell,s}=E_s[\frac{1}{N_{s+T}}\mid N_T=\ell]=\sum_{k=0}^{\ell-1}\frac{1}{1+k}\binom{\ell-1}{k}{}_sp_{x_1}^k(1-{}_sp_{x_1})^{\ell-1-k}
$$
and
$$
\hat b_\ell=E_s[Y,N_T=\ell]=\int_0^Te^{r(T-t)}E_s[N_t,N_T=\ell]\,dt.
$$

\vspace{0.1in}

Then, some elementary binomial manipulations show that
$$
E_s[N_t,N_T=\ell]=\binom{n-1}{\ell-1}{}_Tp_{x_0}^{\ell-1}(1-{}_Tp_{x_0})^{n-\ell-1}[\ell+(n-\ell){}_tp_{x_0}-n\cdot{}_Tp_{x_0}].
$$
Therefore
$$
\hat b_\ell=\binom{n-1}{\ell-1}{}_Tp_{x_0}^{\ell-1}(1-{}_Tp_{x_0})^{n-\ell-1}
\Big[(\ell-n\cdot {}_Tp_{x_0})\frac{e^{rT}-1}{r}+(n-\ell)e^{rT}\int_0^T e^{-rt}{}_tp_{x_0}\,dt\Big].
$$

\vspace{0.1in}

With this, we may compute the corresponding generosity ratios for the ELRiS. We take the pool with $\gamma=2$, other parameters as before, and $m=70$ (resp. 80 and 90). These then imply ratios of  
2.57 (resp.  2.67 and  2.78). In other words, the generosity of the ELRiS plan is not subject to nearly as steep a drop-off as that CPP suffers, with declining modal values of longevity $m$. For comparison, we also include the ratio with $\gamma=10$ and $m=70$, which is 
2.45. That, recall, is a more conservative and risk-averse drawdown plan. It does lessen generosity, but not by a substantial amount.


\endgroup

\newpage
\begin{table}[ht]
    \caption{First Nations Member who Contribute to \& Participate in CPP} 
    \label{tab:mydata}       
    \centering
\includegraphics[width=0.95\textwidth]{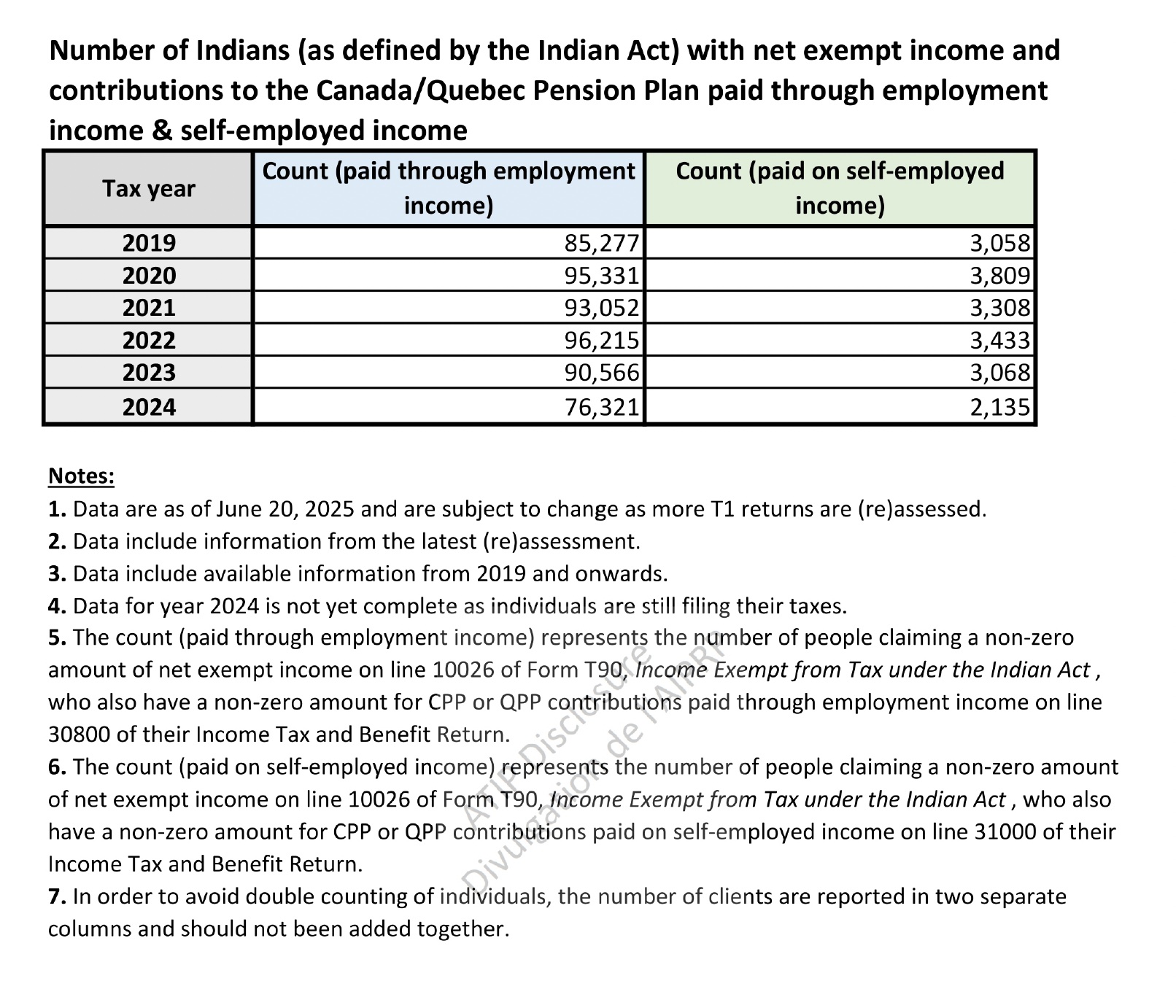} 
\end{table}

\newpage
\begin{table}[ht]
    \caption{Certainty Equivalent Contribution Rate by Employee} 
    \label{tab:mytable}       
    \centering
\includegraphics[width=0.95\textwidth]{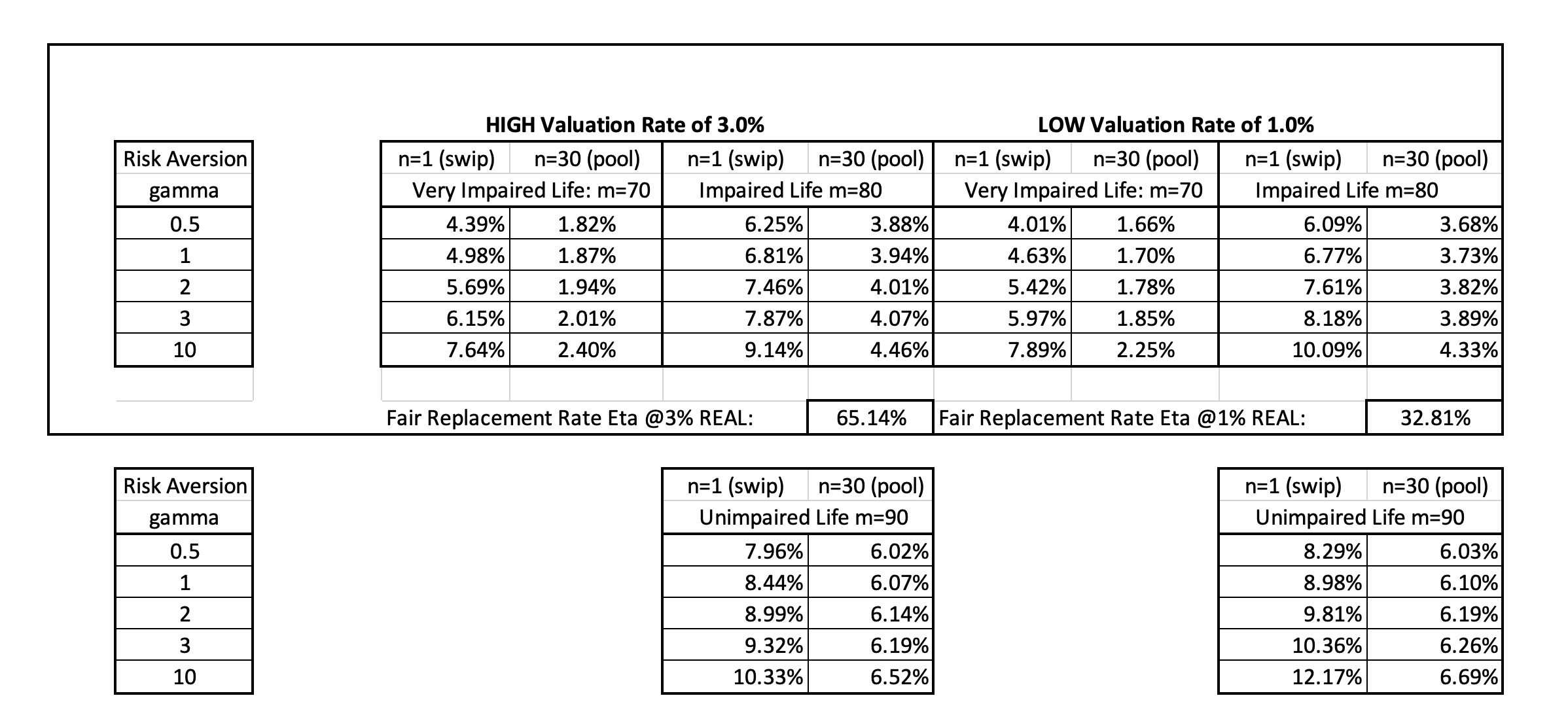} 
\end{table}
\begin{singlespace}
Notes: The table displays the utility-equivalent {\em contribution rate} denoted by $\bar{\alpha}$, which is the percent of working wages that an impaired life ($\bar{m} \leq m$) could contribute to an {\em equitable longevity risk sharing} (ELRiS) plan, and still obtain the same level of discounted lifetime utility. These numbers assume that the employer (or someone) matches the contribution rate, as is the current practice with CPP. The self-employed should double this number.
\end{singlespace}

\clearpage
\begin{singlespace}
Notes: Displays the relative mortality of First Nations peoples of Canada, both on-reserve and off-reserve, across the entire age curve, per Statistics Canada. While infant mortality rates are abnormally high, by a factor of 3 to as high as 5, even at adult ages, the ratio is above one. This also means that life expectancy is lower, regardless of the baseline age. 
\end{singlespace}

\clearpage
\begin{singlespace}
Notes: Simulated paths of yearly income, normalized to a pensionable income of $c=1$, for an {\em equitable longevity risk sharing} (ELRiS) account or plan, with $n=30$ (left) and $n=150$ (right). The plan is optimized assuming a high risk aversion of $\gamma=10$, and uses the calculated contribution rate $\bar\alpha=0.0433$. As a reminder, the conditional probability of an impaired life ($m=80, b=10$) member surviving to age 95, conditional on retiring at age 65, is 1.4\%, which means that the odds of getting to (what we call) ``the wiggle'', is small.
\end{singlespace}

\clearpage
\begin{singlespace}
Notes: Displays the 20'th, 40'th, 60'th and 80'th percentiles of yearly income during retirement, normalized to a pensionable income of $c=1$, for an {\em equitable longevity risk sharing} (ELRiS) account or plan, with $n=30$ (left) and $n=150$ (right). Once again, the plan is optimized assuming a high risk aversion of $\gamma=10$, and uses the calculated contribution rate $\bar\alpha=0.0433$.
\end{singlespace}

\clearpage
\begin{figure}
\vspace{-0.1in}
\caption{Equivalent Pension Contribution Rates (all $\bar m$)}
\label{fig4}
\vspace{-0.5in}
\begin{center}
\includegraphics[width=0.95\textwidth]{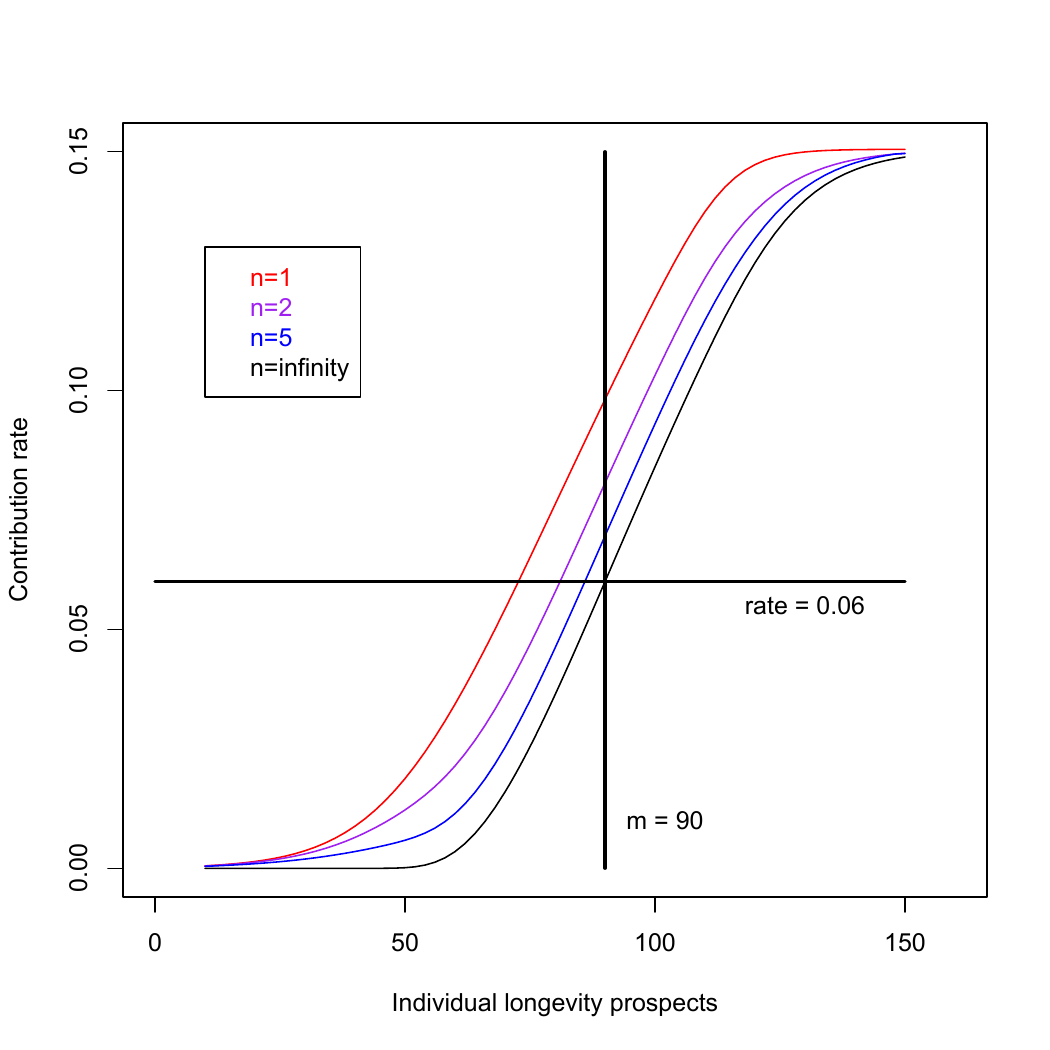} 
\end{center}
\end{figure}
\begin{singlespace}
Notes: This plots the individual contribution rate $\bar\alpha$ (which then must be matched by the employer) against the impaired value of $\bar m$, assuming $r=1\%$, and optimized for $\gamma=2$. We show various (tiny) pool sizes $n$, and the limiting case when $n \rightarrow \infty$.
\end{singlespace}

\clearpage
\begin{figure}
\vspace{-0.1in}
\caption{Equivalent Pension Contribution Rates (realistic $\bar m$)}
\label{fig5}
\vspace{-0.5in}
\begin{center}
\includegraphics[width=0.95\textwidth]{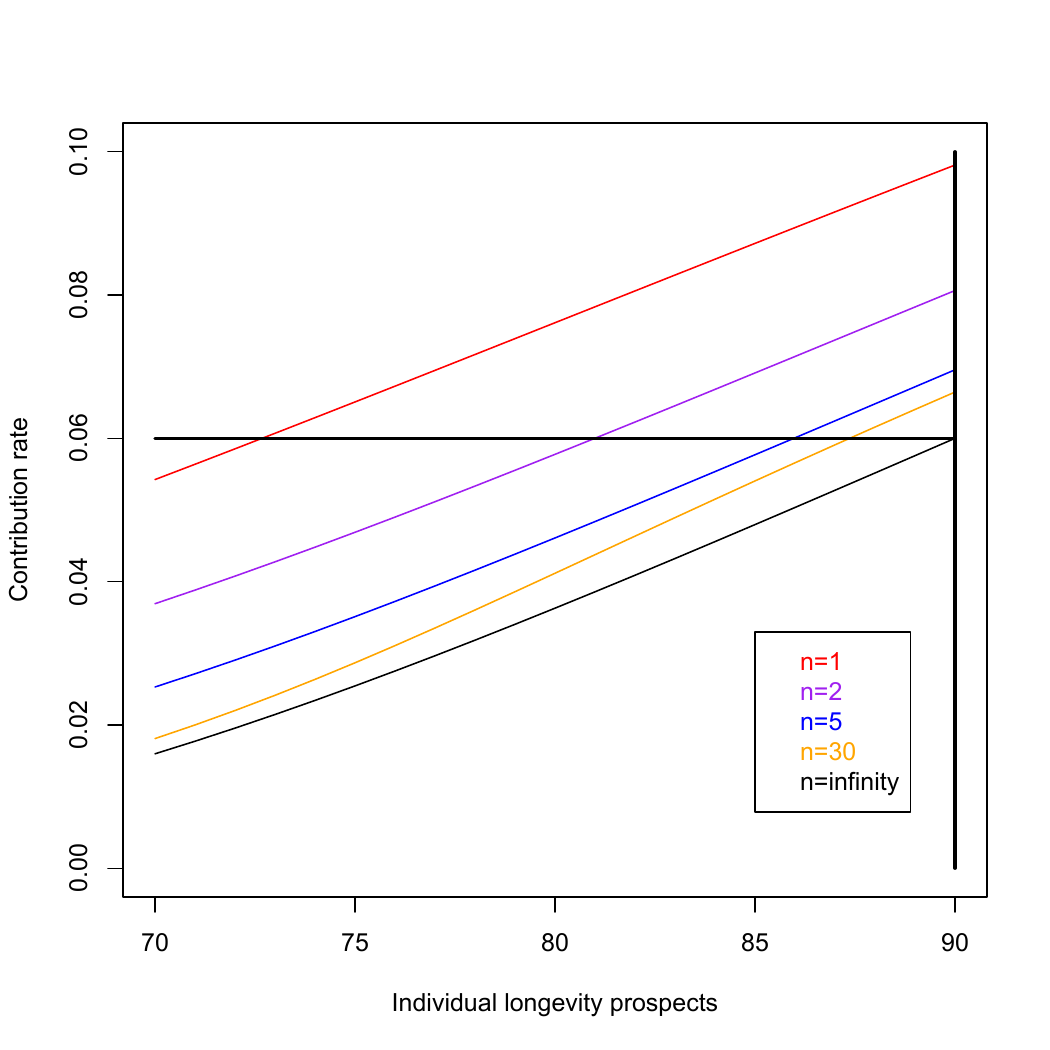} 
\end{center}
\end{figure}
\begin{singlespace}
Notes: This zooms Figure \#\ref{fig4} in on the values of most interest, and adds the $n=30$ case, between 5 and $\infty$. As the modal value of life $m$ increases towards the population's 90 years, and the size of the pool goes to infinity, the individual contribution rate converges to 6\%, which is the same as the CPP.
\end{singlespace}

\end{document}